\documentclass[aps,showpacs,showkeys,preprint,nofootinbib]{revtex4}
\begin{document}
\newcommand{\uG}{\Gamma_{\uparrow}}
\newcommand{\dG}{\Gamma_{\downarrow}}
\newcommand{\dg}{g_{\downarrow}}
\newcommand{\ug}{g_{\uparrow}}
\newcommand{\ab}{\theta}
\bibliographystyle{unsrt}
\title{On a naive construction of kinetic equation}
\author{F. \v{S}anda}
\affiliation {Institute of Physics of Charles University, Faculty
of Mathematics and Physics, \\Ke Karlovu 5, 121 16 Prague 2, Czech
Republic\\(Tel. (**-420-2)2191-1341, Fax (**-420-2)2492-2797,
E-mail sanda@karlov.mff.cuni.cz)}
\date{\today}
\begin{abstract}
We return to the subject of stability of infinite time asymptotics
of kinetic equations. We found a model which is simpler than those
studied previously and which shows unstable behavior corresponding
to our arguments from \cite{moje}, where, however, a relatively
complicated problem was treated. Our simplification to four levels
interacting with surroundings enable us to proceed easily through
all the way with just a pen and paper. We provide no numerical
modelling whose justification causes naturally difficulties to the
reader. We draw also further consequences of the found
instability, not only with respect to higher order terms in
kinetic equations but also concerning the very philosophy of
physical modelling. The latter point can give more practically
oriented physicist even better motivation than mere speculations
about potential instabilities due to higher order terms in
perturbation treatments without concrete resolution of correct
asymptotics.
\end{abstract}
\maketitle
\section{Introduction} Subject of our paper are new
results questioning stability of solution of kinetic equations.
Kinetic equations are widely used in solid state or statistical
physics in modelling of transfer processes including relaxation
phenomena, influence of external fields etc. This concerns a great
number of physical theories appropriate for different physical
regimes of interest like Boltzmann equation \cite{boltzmann},
Fokker-Planck equation \cite{focker}, Pauli master equation
\cite{pauli}, or its generalized version introduced independently
in different forms by Zwanzig, Mori \cite{Zwan,ToMo} etc. General
mathematical structure of these theories is the set of
differential or integro-differential equations (of the first
order), which determine time evolution of quantities of interest.
Specific type of these differential equations is not unique, in
its easiest form we meet with time independent Markov process
(without memory), the other may contain time dependencies in
coefficients (for example dependence on external fields), memory
terms (time nonlocal equations), or inhomogenous terms or
nonlinearity. Instabilities or chaotic behavior in case of
complicated nonlinear equations are surely not very surprising.
Below we will deal with the easiest form of such equations - the
set of linear differential equations with constant coefficients.
Specific topic of this work is scrutiny of stability of steady
state, including that of the attractor nature of the steady state.
Let us give some comments here, how this treatment is related to
the other types of mathematical structure which are also used in
physically similar treatments, and about physical consequences of
this work. Time-local and memory-including theory of Markov
processes are related in general (for general mathematical theory-
see for example \cite{mark}) and also explicit equivalence for
specific type of equations stemming from the Liouville equation
could be found \cite{mark1}.
 The goal of this work is to argue that memory for
Markov processes can be integrated into time local coefficients.
In addition we are interested in steady state and infinite time
asymptotics. Coefficients (in evolution matrix) often turn into
time independent ones in infinite time region \cite{mark2,mark3}.
Some kind of Markov approximation here becomes exact. This
provides the connection with our work. Influence of external
fields is further property which can be often related with topic
of our interest.

The origin of kinetic equations is mostly in perturbational
theories, which turn  fundamental microscopical physical laws into
differential equations determining time evolution of macroscopic
quantities available in experiments. For practical purposes the
coefficients are usually  approximated by calculation  of leading
terms of the Taylor series in a perturbational parameter. Quite
usual is using approximation of the second order in coupling to
uncontrolled degrees of freedom, what enables to incorporate
connected relaxation phenomena.

Then the time evolution of "quantities of physical interest" is
solved. Formally, validity of all these approximations of the
perturbational origin is limited on the time axis. On the other
hand, these are the steady state and asymptotic limit which are of
the greatest physical importance. Otherwise well applicable theory
is thus asked to achieve reasonable results also for this time
region. Expected results like the Boltzmann distribution were
found in the simplest kinetic models, what gave physicists strong
belief in general applicability of the particular kinetic theory.
There is a statement that little changes of coefficients can not
change in very dramatic way results of numerical studies, which is
usually considered as "physically reasonable" . However, we want
to argue that this statement is {\bf NOT} true. Maybe some people
(mainly mathematically oriented) can justify such a statement in
general, but they mostly think, that it is the matter of purely
mathematical toys that are seldom present in normal day physics.
Our warning is: Nothing is so far from the truth.

Recently we found in \cite{moje} rather unexpected instability in
asymptotical behavior in case of a not very complicated open
system. The latter showed some not very standard features, but no
direct indication of internal collapse in modelling of time
development, of the type of, e.g., a clearly unphysical result.
Many model approximations, where one can meet the below described
instability, may seem to be quite usual or not very suspicious.

However, whatever is the truth about correct time asymptotics of
the model hamiltonian from \cite{moje}, we would like to warn the
reader against unexpected features encountered in this problem.
When one incorporates (according to his/her opinion) important
physical processes into kinetic equations, attention must be also
focused on stability of the solution of the problem. In particular
the question is which part of the results is correct and which is
only a belief (and contingently only an unjustified belief). In
case when  the resulted steady state can not be confronted with,
e.g., thermodynamical laws, the result of thus invalid simulation
may seem to be quite good.

 Sometimes physical belief in the second order
approximation results is hidden into sophisticated mathematical
methods which pretend to give a "general" proof of correctness of
the second order results without realistic treatment of the
stability, case by case. Such methods might, regardless of their
mathematical validity and usefulness, in concrete applications
leave a space for speculations about existence of instabilities,
in some important features of the solution, like in our model from
\cite{moje}. In \cite{moje} we showed how it is possible to fulfil
so called Davies theorems \cite{Dav} and, simultaneously, also
obtain sharp instability in the long time regime behavior against
higher order contributions to Tokuyama-Mori coefficients. The
proper quantity of our interest is time asymptotics of the kinetic
equation (regardless of whether the steady state is given only by
internal dynamics, or it is influenced by some regular external
field, presence of interchange particle etc.).

Organization of our paper is the following: In part 2 we introduce
a quite usual transfer process model - four site open quantum
system interacting with surroundings. We perform standard
treatment using second order perturbational theory of relaxation
 with full spectral analysis of evolution matrix and the unique asymptotic
 steady state will be found. One can verify that the
 model, at least in standard thinking, contains description of all
 the important physical processes present here.
 Further we introduce (in section 3) a correction
 to evolution matrix. We thus add a transfer channel, whose description
 is constructed in the same way like that of the previous ones.
 However, we assume the effectiveness of the new channel to be very small
 with respect to that of the original channels.
 This new process is also not apparently seen to change overall
 properties of system. Nevertheless,
 dramatic changes in the steady state will appear.
 The small perturbation will be treated in two distinct ways.
First we consider it like a higher order correction in the
perturbational series to our second order approximation. This
approach is in our view mathematically more interesting, but is
rather speculative because we do not explicitly calculate all the
fourth order contribution that can come from higher order
calculation in some exact microscopical theory. However, the
instability of the second order approximation is verified
regardless this objection. The second possibility is to complement
the model by a further small parameter which will be limited to
zero after the asymptotic calculation. In this case we fully stay
at a position of the standard second order kinetic theory, but we
consider model character of our hamiltonian as in the everyday
physics, and we treat stability of the model conclusions against
"omitted physically irrelevant" process. In part 4 we throw some
light on indication of instability of the given type. At first we
show its indication in spectrum of the evolution matrix, then a
purely analytical treatment of stability will be given. In part 5
we compare our result with van Hove limit of the same model
Hamiltonian. We will find a connection with some pathological case
of this well understood limit, where kinetic theory also may have
problems with establishing of asymptotical state.

 All these ideas
are presented on this simple model. Further consequences about
relation to Davies theorems are not repeated , interested reader
can find them in \cite{moje}. All the calculations are hand made
product and the reader is pretty invited to follow them. The
reader is also invited to think how the referred problems, which
one may consider as clear or quite trivial, may become forgotten
when a complicated model is treated which is not analytically
calculable, but where only computer simulations are at hand.
\section{Model}
Let us consider the following model hamiltonian of a four site
open system. We measure energy in units  $\hbar$.
\begin{equation}
 H= \epsilon (c_1^{\dagger}c_1+c_3^{\dagger}c_3) + J
 (c_2^{\dagger}c_3+c_3^{\dagger}c_2)+
\sum_k \{G^{(1-2)}_k (B_k c_1^{\dagger}c_2 + B^{\dagger}_k
c_2^{\dagger}c_1) + G^{(3-4)}_k (B_k c_3^{\dagger}c_4 +
B^{\dagger}_k c_4^{\dagger}c_3) + \Omega_k B^{\dagger}_k B_k.\}
\label{Ham}\end{equation} This hamiltonian describes our four site
system where creation $c^{\dagger}_i$ and annihilation operators
$c_i$ are each related to the i-th site, with three transfer
channels, two of them being bath-induced the remaining one being
coherent. As far as we consider one particle only, there is no
necessity to introduce (anti)commutational relations between these
operators. Dynamics of the system is dominated by bath induced
transfer channels between 1-2 and 3-4 sites and a coherent
transfer channel 2-3. $B^{\dagger}_k, B_k$ are creation and
annihilation operators of k-th bath phonon mode ( fulfilling boson
commutational relation), $G^{(i-j)}_k$ are coupling constants of
the system-bath interaction. Parameter J describes power of the
coherent channel.

We do not allow the interference between bath induced channels 1-2
and 3-4, what means to fulfil conditions
like:\[\sum_k\delta(\epsilon - \Omega_k) G^{(1-2)*}_k G^{(3-4)}_k
Tr_{Bath}( \rho_{Bath} B^{\dagger}_k B_k) =0.
\] This can be generally fulfilled if one considers that the
particular transfer channels are induced by different phonon
modes\[G^{(1-2)}_k G^{(3-4)}_k =0.\]

One can treat this hamiltonian using various schemes. Firstly, it
is possible to think about different regimes, according to
different magnitude of the coefficients in Hamiltonian. We  are
here interested in the regime where it is appropriate to treat J-
and bath-induced transfer channel as a perturbation. We emphasize
that this choice does not correspond to so-called van Hove limit
\cite{vHove}. One can diversify the physical interpretation of
hamiltonian (\ref{Ham}) and a chosen perturbation scheme. Coherent
channel one can consider as a slow internal motion treated
according to \cite{capbar}, but it may also represent constant or
periodical external field influence.

Various constructions of kinetic equations can be also applied.
 We restrict ourselves to those which respect chosen mathematical
structure and the physical regime. Though also here physicist use
various formalisms, one may obtain our results using
Nakajima-Zwanzig identity, Tokuyama-Mori equation (both in their
second order approximation), and also Haken-Strobl-Reineker
parameterization \cite{HSR1,HSR2}; all these ways lead to formally
the same master equation:
\begin{equation}
\frac{d \rho_{ij}}{dt}=\sum_{\{kl\}} W_{\{ij\},\{kl\}}
\rho_{\{kl\}} \label{kinetic}
\end{equation}

where vector $\rho$ is organized in the following way \[
\rho^T = (\begin{array}{ccccccccc}\rho_{11}, & \rho_{22}, &
\rho_{33}, & \rho_{44}, &  Re\rho_{23}, & Im\rho_{23}, &
Re\rho_{12}, & Im\rho_{12},& Re \rho_{13},\end{array} \]
\[\begin{array}{ccccccc}
Im\rho_{13}, & Re\rho_{34}, & Im\rho_{34}, & Re\rho_{24}, &
Im\rho_{24}, & Re\rho_{14}, & Im\rho_{14}
\end{array} )
\]
Matrix $W$ we call evolution matrix. $W^{(2)}$ is the second order
approximation of $W$. It reads
\begin{equation}
W^{(2)} =\left(\begin{array}{cccc}  A & 0 & 0 & 0\\ 0 & B & 0& 0
\\ 0 & 0 & C & 0 \\ 0 & 0 &  0 & D
\end{array}\right)
\label{approx2}
\end{equation}

where
\begin{equation}
A=\left(\begin{array}{cccccc} - \dG & \uG & 0& 0& 0 & 0
\\ \dG & -\uG & 0 & 0 & 0 & -2J \\ 0 & 0& -\dG & \uG & 0& 2J\\
0 & 0 &\dG & -\uG & 0 & 0 \\ 0 & 0 & 0 & 0 & -\frac{\uG + \dG}{2}&
- \epsilon \\ 0 & J & -J & 0 & \epsilon &-\frac{\uG + \dG}{2}
\end{array}\right),
\end{equation}
\begin{equation}
B=\left(\begin{array}{cccc}-\frac{\uG + \dG }{2}& \epsilon & 0 &
-J \\ -\epsilon &-\frac{\uG + \dG }{2} & J & 0 \\ 0 & -J & -\dG
 & 0 \\ J & 0 & 0 & -\dG
\end{array}\right),
\end{equation}
\begin{equation}
C=\left(\begin{array}{cccc}-\frac{\uG + \dG }{2}& \epsilon & 0 & J
\\ -\epsilon &-\frac{\uG + \dG }{2} & -J & 0 \\ 0 & J & -\uG
 & 0 \\ -J & 0 & 0 & -\uG
\end{array}\right),
\end{equation}
\begin{equation}
D=\left(\begin{array}{cc}-\frac{\uG+\dG}{2} & \epsilon \\
-\epsilon & -\frac{\uG+\dG}{2}
\end{array}\right).
\end{equation}
Here\[\uG=2\pi\sum_k [G^{(1-2)}_k]^2 \delta (\epsilon-
\Omega_k)Tr_{Bath} \rho_{Bath} (B^{\dagger}_k B_k) =2\pi\sum_k
[G^{(3-4)}_k]^2 \delta (\epsilon- \Omega_k)Tr_{Bath}( \rho_{Bath}
B^{\dagger}_k B_k)\]
\begin{equation} \dG=2\pi\sum_k [G^{(1-2)}_k]^2
\delta (\epsilon- \Omega_k)Tr_{Bath}( \rho_{Bath} B_k
B^{\dagger}_k)
 =2\pi\sum_k [G^{(3-4)}_k]^2 \delta (\epsilon- \Omega_k)Tr_{Bath}
(\rho_{Bath} B_k B^{\dagger}_k). \end{equation} The equality of
coefficients for 1-2 and 3-4 transfer is our additional
assumption, that can not be deduced from (\ref{Ham}). Notice that
$J, \uG, \dG$ we consider as perturbations of the same magnitude,
proportional to the parameter $\lambda^2$.
\begin{equation}
J,\uG, \dG \quad \propto  \lambda^2\label{schema0}
\end{equation}
(Reason the proportionality is only a consistency with standard
perturbational order of bath-induced transfer channel.)

The steady state is given by the condition:
\begin{equation}
\sum_{\{kl\}} W_{\{ij\},\{kl\}} \rho_{\{kl\}}=0
\label{steady1}\end{equation}
 We are now to calculate the
complete spectrum of the evolution matrix. Firstly: this enables
us to show that steady state is also the unique asymptotic state
of this equation. Furthermore we will argue that the solution has
no apparent deviant feature. Last but not least, in section 4 we
will show that in a careful treatment one can indicate, in this
spectrum, the instability calculated below.

The evolution matrix was arranged so that it has a quasidiagonal
structure. We have to calculate a characteristic equation. After
bit of algebra (we must calculate determinant of submatrices of
maximal order 6) and rearranging resulting terms, we obtain
\footnote{The terms are ordered according to ordering of
submatrices; one submatrix is one row.}:
\[
0=\xi\cdot(\xi+\uG+\dG)\cdot
\{\xi(\xi+\uG+\dG)[(\xi+\frac{\uG+\dG}{2})^2+\epsilon^2]+4J^2(\xi+\frac{\uG+\dG}{2})^2\}
\]
\[\cdot \{\xi+i\epsilon+\frac{\uG+\dG}{2})(\xi+\dG)+J^2\}\cdot
\{\xi-i\epsilon+\frac{\uG+\dG}{2})(\xi+\dG)+J^2\}\]
\[\cdot \{\xi+i\epsilon+\frac{\uG+\dG}{2})(\xi+\uG)+J^2\}\cdot
\{\xi-i\epsilon+\frac{\uG+\dG}{2})(\xi+\uG)+J^2\}\]
\begin{equation}
\cdot[(\xi+\frac{\uG+\dG}{2})^2 +\epsilon^2]
\end{equation} Twelve roots can be calculated directly from the quadratic terms.
What remains is an equation of the fourth order. The roots can be
in principe also extracted using Cardano formula, but it does not
provide an easy survey. Instead we inspect behavior in the
$\lambda \rightarrow 0$ limit of the perturbational parameter.
This analysis and calculation of twelve exact eigenvectors is
provided in Appendix.

In conclusion: there is only one steady state and, at least for
not very high parameter $\lambda$, all the other eigenvectors of
matrix \ref{approx2} have negative real parts, i.e. connected
terms in time evolution simulation disappear in the infinite time
and the steady state is also the asymptotical one. Because of
finite order of the matrix there is a region surrounding the zero,
where this problem has infinite time asymptotics given by
(\ref{steady1}), so one can limit himself/herself to this region
without complications. For very high values of $\lambda$
parameter, the model need not have the correct behavior in
accordance with general inapplicability of the perturbational
treatment for this case. The worth notice is that the zero
eigenvalue came purely from the first submatrix A. The others have
nonzero determinants and thus the only solution of the steady
state condition  must be zero for associated elements in the
density matrix.

 We can give at this place the solution to steady state
 condition (\ref{steady1}). We work with normalization condition
 \begin{equation}
 \sum_{i}\rho_{ii}=1.
 \label{norm}
\end{equation}
Then the result is:\begin{equation}
\rho_{11}=\frac{\uG^2}{(\uG+\dG)^2};\quad
\rho_{22}=\rho_{33}=\frac{\uG \dG }{(\uG+\dG)^2};\quad
\rho_{44}=\frac{\dG^2}{(\uG+\dG)^2};\quad \rho_{i \neq j}=0.
\label{result2}\end{equation} We specifically note the equality in
population at sites 2 and 3. The reader may speculate whether this
model and result are in whatever sense bad. In any case, there is
no internal collapse in these calculations. One may be suspicious
about the fact that the model does not lead to thermodynamical
equilibrium, but as we notice above we do not argue that this
system is in thermodynamical equilibrium as we are not in the van
Hove limit, and also the purely internal character of the coherent
transfer was not specified. There are some physicists who believe
that the validity of thermodynamical laws must be in some
direction connected with the van Hove limit \cite{cap4}. The 2-3
channel is elastic what implies 2-3 symmetry and consequent
equality $\rho_{22}=\rho_{33}$. The transfer term proportional to
$J$ can come also from defined (e.g. harmonic) external field;
then thermodynamical prediction can fail or this prediction need
not be clear without further calculations.

\section{Perturbation} In this section we introduce a small perturbation of
model (\ref{Ham}) in form of an incoherent transfer channel
between sites 2 and 3. The new terms in evolution matrix can be
quite small with respect to the other ones coming from the
previous consideration. Construction of the terms is fully
analogical to the previous one. Formally one can include a change
into hamiltonian :\[\delta H = \sum_k G^{(2-3)}_k (B_k
c_3^{\dagger}c_2 + B_k^{\dagger}c_2^{\dagger}c_3).\] We refer a
new corrected evolution matrix, the effectiveness of the new 2-3
channel is measured by rate coefficients $\ug, \dg$:

 \begin{equation} A=\left(\begin{array}{cccccc} - \dG
& \uG & 0& 0& 0 & 0
\\ \dG & -\uG -\ug & \dg & 0 & 0 & -2J \\ 0 & \ug & -\dG -\dg & \uG & 0& 2J\\
0 & 0 &\dG & -\uG & 0 & 0 \\ 0 & 0 & 0 & 0 & -\frac{\uG + \dG +
\ug + \dg }{2}& - \epsilon \\ 0 & J & -J & 0 & \epsilon
&-\frac{\uG + \dG + \ug + \dg}{2},
\end{array}\right)
\end{equation}
\begin{equation}
B=\left(\begin{array}{cccc}-\frac{\uG + \dG + \ug}{2}& \epsilon &
0 & -J \\ -\epsilon &-\frac{\uG + \dG + \ug}{2} & J & 0 \\ 0 & -J
& -\dG -\frac{\dg}{2} & 0 \\ J & 0 & 0 & -\dG -\frac{\dg}{2}
\end{array}\right),
\end{equation}
\begin{equation}
C=\left(\begin{array}{cccc}-\frac{\uG + \dG + \dg}{2}& \epsilon &
0 & J \\ -\epsilon &-\frac{\uG + \dG + \dg}{2} & - J & 0 \\ 0 & J
& -\uG -\frac{\ug}{2} & 0 \\ -J & 0 & 0 & -\uG -\frac{\ug}{2}
\end{array}\right),
\quad
D=\left(\begin{array}{cc}-\frac{\uG+\dG}{2} & \epsilon \\
-\epsilon & -\frac{\uG+\dG}{2}. \label{approx4}
\end{array}\right)
\end{equation}
Now we calculate the stationary state of problem (\ref{approx4}).
We again work with the assumption that $\lambda$ is so small that
submatrices B, C and D are regular and so the stationary condition
applied here has the trivial solution only. Then one gets the
result after an easy algebra:\[
\rho_{11}=C\uG^2\{\dg(\frac{2\epsilon^2}{\uG+\dG+\ug+\dg}+\frac{\uG+\dG+\ug+\dg}{2})+2J^2\},
 \]
 \[
 \rho_{22}=C\dG\uG\{\dg(\frac{2\epsilon^2}{\uG+\dG+\ug+\dg}+\frac{\uG+\dG+\ug+\dg}{2})+2J^2\},\]
   \[\rho_{33}=C\dG\uG\{\ug(\frac{2\epsilon^2}{\uG+\dG+\ug+\dg}+\frac{\uG+\dG+\ug+\dg}{2})+2J^2\},\]
 \[\rho_{44}=C\dG^2\{\ug(\frac{2\epsilon^2}{\uG+\dG+\ug+\dg}+\frac{\uG+\dG+\ug+\dg}{2})+2J^2\},
 \]
\[Re \rho_{23}=C\frac{-2\epsilon J(\dg-\ug)\dG\uG}{(\uG+\dG+\ug+\dg)},\]
\[ Im \rho_{23}=CJ(\dg-\ug)\dG\uG\]
where $C$ is a normalization constant to be deduced from
(\ref{norm}):\[ \frac{1}{C}=
2*J^2(\uG+\dG)^2+(\frac{2\epsilon^2}{\uG+\dG+\ug+\dg}+\frac{\uG+\dG+\ug+\dg}{2})\cdot
(\uG+\dG)(\uG\dg+\dG\ug)
\]

We are interested especially in the ratio of the population on the
sites 2 and 3. The reason for this specific interest becomes
apparent later.
\begin{equation}\frac{\rho_{22}}{\rho_{33}}=\frac{\dg(\frac{2\epsilon^2}{\uG+\dG+\ug+\dg}+\frac{\uG+\dG+\ug+\dg}{2})+2J^2}{
\ug(\frac{2\epsilon^2}{\uG+\dG+\ug+\dg}+\frac{\uG+\dG+\ug+\dg}{2})+2J^2}
\label{result4}\end{equation} The term "small perturbation" has to
be formalized in order to talk about the instability. We have
worked out this point in two different ways. First of them is
submitted mainly for a mathematically oriented reader. We consider
$\ug,\dg$ as proportional to $\lambda^4$.
\begin{equation} J,\dG,\uG \propto \lambda^2 \quad \ug, \dg \propto \lambda^4
 \label{schema1}\end{equation}
One may have some physical objections against this interpretation,
stemming from the fact, that we did not provide complete 4-th
order inspection of the kinetic theory. But we have quite narrow
ambition here. We point out the instability of the result
(\ref{result2}) against the 4-th order correction, which we
consider to be arbitrary - as a potentiality. We argue that an
arbitrary perturbation could be used to achieve this conclusion.
The physical motivation in this interpretation is in the
background only, in order to get the reader interested, the
statement is of mathematical character. One can also omit here the
additional term in Hamiltonian $\delta H$ , and think of the
perturbation  as of a higher order terms coming potentially from
the Hamiltonian (\ref{Ham}) that are omitted in standard second
order calculation. We will compare the results (\ref{result2}) and
(\ref{approx4}) in $\lambda \rightarrow 0$ limit where the
perturbational treatment is best verified. (Performing this limit
has of course no consequence in connection with the main statement
- instability.)

 The second interpretation stays fully on the position of the second order
 kinetic equation.  We introduce some further parameter $\eta$  that measures relative power of different transfer channel
  \begin{equation}J,\dG,\uG \propto \lambda^2 \quad \ug,\dg \propto \eta
  \lambda^2.
  \label{schema2} \end{equation}
After evaluation of the $\lambda \rightarrow 0$ limit, which
 gives precise mathematical sense to our calculation, we consider
 $\eta$ to be small, formally limiting to 0.
 We shall show that  regardless the arbitrarily small (but nonzero) magnitude of $\eta$,
 the result (\ref{result2})
 is not preserved. In other words, \[\lim _{\eta\rightarrow
 0}\rho(\eta) \neq \rho(\eta =0) \] $\rho$ designates here the steady
 state in $\lambda\rightarrow 0$ limit.
This is the central statement that we are going to prove.
  We inspect the
 stability of standard kinetic equations with respect to physical
 processes which were not incorporated into a model in question because of
their not high strength (at least from a formal, cursory point of
view) and consequent underestimating their influence. This point
is possibly not so interesting mathematically because the second
order theory is held here, but it seriously questions the
straightforward applicability of the standard kinetics from the
physical point of view. Both these interpretations are from
mathematical and also from physical context quite distinct. We
argue that the instable behavior is the internal problem of the
approximation (\ref{approx2}) and does not come from the very
specialized choice of perturbation or scheme of its treatment. In
the next subsection we make this point more clear. \newline
Performing the announced limits: \newline First interpretation
(\ref{schema1}):
\[\lim_{\lambda\rightarrow 0 }
\frac{\rho_{22}}{\rho_{33}}=\lim_{\lambda\rightarrow 0
}\frac{\lambda^4\dg(\frac{2\epsilon^2}{\lambda^2(\uG+\dG)+\lambda^4(\ug+\dg)}+
\frac{\lambda^2(\uG+\dG)+\lambda^4(\ug+\dg)}{2})+2\lambda^4J^2}{
\lambda^4\ug(\frac{2\epsilon^2}{\lambda^2(\uG+\dG)+\lambda^4(\ug+\dg)}+\frac{\lambda^2(\uG+\dG)+\lambda^4(\ug+\dg)}{2})+2\lambda^4
J^2}= \frac{\dg}{\ug}\] Second interpretation (\ref{schema2}):
\[\lim_{\eta\rightarrow 0} \lim_{\lambda\rightarrow
0}\frac{\rho_{22}}{\rho_{33}}=\lim_{\eta\rightarrow
0}\lim_{\lambda\rightarrow 0
}\frac{\eta\lambda^2\dg(\frac{2\epsilon^2}{\lambda^2[\uG+\dG+\eta(\ug+\dg)]}+
\frac{\lambda^2[\uG+\dG+\eta(\ug+\dg)]}{2})+2\lambda^4J^2}{
\eta\lambda^2\ug(\frac{2\epsilon^2}{\lambda^2[\uG+\dG+\eta(\ug+\dg)]}+\frac{\lambda^2[\uG+\dG+\eta(\ug+\dg)]}{2})+2\lambda^4
J^2}= \lim_{\eta \rightarrow 0}\frac{\dg}{\ug}=\frac{\dg}{\ug}\]

Also other results are identical in both our limits, we refer it
in a short way.

\[
\rho_{11}=\frac{\uG^2\dg}{(\uG+\dG)(\uG\dg+\dG\ug)}
 \]\[  \rho_{22}=\frac{\dG\uG\dg}{(\uG+\dG)(\uG\dg+\dG\ug)}\]
   \[\rho_{33}=\frac{\dG\uG\ug}{(\uG+\dG)(\uG\dg+\dG\ug)}\]
 \[\rho_{44}=\frac{\dG^2\ug}{(\uG+\dG)(\uG\dg+\dG\ug)}
 \]
\begin{equation}Re \rho_{23}=0 ; \quad  Im \rho_{23}=0\end{equation}

We clearly see that if $\ug \neq \dg$, then sharp changes appear
in solution (\ref{result2}). We did not assume equality between
these coefficients. Moreover the physical motivation of our
correction points out to $\ug \neq \dg$. Rather the standard
thermodynamics relation
\[ \frac{\ug}{\dg}=\exp{\beta \epsilon}
\]
may be assumed. Derivation of this statement consists in some
additional assumptions about initial state of the bath, which are,
however, standard. Of course, if one is interested in the Taylor
series structure in higher order expansion of (\ref{result4}), our
interpretations are mutually different, but the general picture is
not changed.

We would like to give further warning here. One need not be very
surprised because of the following argumentation. Transfer channel
connected with parameter J is in fact physically  also of the
fourth order, because its direct application to density matrix
(commutator $[J,\rho ]$ ) change either the ket or bra side of
density matrix only and comparable process connecting the diagonal
terms in density matrix is thus of the fourth order. Then this new
included channel is comparably strong (in the first
interpretation) or even stronger (in the second argumentation)
than the first one. We give threefold
counterargument:\begin{description}
\item[1,] - Nevertheless, formally the coherent process (J-proportional) is included in the second
order. Such a treatment is absolutely standard. Care in this
direction becomes difficult or technically unable for complicated
system. Moreover this objection also seriously questions concept
of the generalized master equation in general, because treatment
of the whole density matrix (of the system) included information
about set of generally incompatible observables. With respect to a
particular measurement (here for example site probability
measurements) the other terms unrelated to this measurement (here
off-diagonal matrix element) have always the role of some kind of
memory. So their treatment apparently differs from that of bath
induced channel one.

\item[2,] - Physical intuition in more complicated
case is uncertain and may fail.

\item[3,] - 
If one is internally sure about his/her intuition, please try
 inspect the formula (\ref{result4}) again now with the following
sheme\[ \uG, \dG, J \propto \lambda^2;\quad \ug ,\dg \propto
\lambda^5
\] or alternatively
\[\uG, \dG, J \propto \lambda^2 \quad  \quad \ug ,\dg \propto \eta \lambda^4
.\] One can see here that in the same limits as above, the
instability is still present, though the perturbation should be
now smaller than the fourth order coherent channel. In fact the
coherent channel transfer is seemingly of the higher order then 4!
Have you seen it before this limit calculation? \end{description}
\section{Indication of instability}
We have argued in the previous section, that instability described
above is the matter of internal problem of approximation
(\ref{approx2}). Despite of its reasonable behavior in time
evolution simulation which we proved in Appendix, a problematic
step has been indicated above and this problem must be visible
purely from the second order approximation calculation. In fact,
if one were not able to give some indication of the instability of
the approximation from itself, the situation would have been
critical. At first: Calculation in higher order is not standard,
and what is worse, it is an extremely difficult task. The second
problem is that in that case one can never ( in no finite order of
calculation) be sure whether the provided approximation is already
stable. We give two points which are connected with our
instability and indicate it.
\subsection{Long-time excitation in the spectrum}
This method enables us to show quite simply how to obtain the
indication purely from spectrum of the evolution matrix in the
second order approximation.  What is important from practical
point of view, one may use this method without principal
difficulties also in complicated models by numerical analysis and,
further, may in many ways also implement orientational indication
into her/his time evolution computer simulation. Only for
simplicity and without any change in physical context we assume
that the number of linearly independent eigenvectors equals to the
order of the matrix. In problem (\ref{approx2}) this statement is
proven, because none of the submatrix has two identical
eigenvalue. The general form of spectral decomposition of finite
(non-normal) matrix is then:
\[W_{ij}= \sum_q \xi_q (L_q)_i  (R_q)_j
\]
where $\xi_q$ is q-th eigenvalue and $R_q (L_q)$ is the associated
right(left i.e usual) row (column) eigenvector. This decomposition
holds good this normalization: \[ \sum_i (R_q)_i (L_q')_i=
\delta_{q q'}.
\]
Kinetic equations conserve total probability, what results into
the fact that eigenvalue 0 is always present. Have a look at our
result in the Appendix again. The very suspicious eigenvalue is
$\xi_4$. In perturbation scheme (\ref{schema0}), there is a
proportionality $\xi_4 \propto \lambda^6$.  No wonder that this
eigenvalue need not be very stable against our perturbation
regardless of the type of scheme. More generally: Let us have, in
our spectrum, eigenvalue with real part approaching zero (with
$\lambda \rightarrow 0$) and proportional to higher than second
power of $\lambda$ ($n
> 2$):
\[W_{ij}=0\cdot (L_0)_i(R_0)_j +a_1\lambda^n\cdot (L_1)_i(R_1)_j +
\ldots\] We explicitly emphasize that also $(L_q),(R_q)$ are
$\lambda$-dependent what only enables that $W_{ij}$ has terms of
just the second order.

 Then one can easily construct the mathematical
"perturbation" which causes the instability, for example
perturbation
\[ \delta W_{ij}=-\lambda^n\cdot (L_0)_i(R_0)_j - a_1\lambda^n\cdot
(L_1)_i(R_1)_j +{\cal O}(\lambda^n)\] that change the stationary
state from $(L_0)$ to $(L_1)$ in the limit.

In the first perturbational technique (\ref{schema1}) possibility
of such construction is straightforward. In the second case
(\ref{schema2}) we must construct it from the more complicated
terms, but one can see in our model (\ref{approx4}) that it is
possible.
 Of course, not each one of the mathematical "perturbations" is
physically interpretable. One usually has some conditions for the
evolution matrix stemming from the conservation laws (at least
particle conservation in case of solid state physics), etc., but
the set of possible perturbations is so great, that it surely
contains also reasonable perturbations. One of them we introduced
in subsection 3. In our problem, there is the "near-the-zero"
eigenvalue proportional to the sixth order in $\lambda$, so we
could choose the perturbation smaller than we have done.
Concluding this subsection we notice that the reciprocal real part
of eigenvalue (without sign) can be also called the lifetime of
"excitation" (though it need not be a very appropriate name in
some cases) . This provides practical indication of this
instability - highly increasing lifetime of relaxation phenomena
when the parameter of perturbation is reduced according to formal
scheme of construction of given kinetic equation. Because of
clarity of this point we do not give any formal statement.

\subsection{Analytical treatment of stationary condition}
In this subsection we give an easy example of analytical
stationary condition (\ref{steady1}) treatment. This treatment is
stable against perturbation. We know that it need not be usually
the very appropriate method from practical point of view. When a
direct explicit resolution is not available (for more complicated
or extended problems) one must take extreme care in computational
implementation about numerical errors. The main reason for
introducing this calculation is further understanding of the
origin of the instability for the reader who still has not
internally accepted the presented facts. For our treatment we need
(for simplicity) to assume that there is a unique asymptotical
state of the system. This is because we will here take care of the
stability of the zero eigenvector of the evolution matrix only.
This treatment does not provide the proof that there is no
potential eigenvalue with positive real part (collapse of model)
or there is some eigenvalue along the imaginary axes so near to
zero (real part) that it can approach through some perturbation
the imaginary axis. We look for resolution of approximation
(\ref{approx2}) as Taylor series coefficients.
\[\rho=\sum_n \lambda^n \rho^{(n)}
\]
The important difference as compared to Taylor series of solution
(\ref{result2}) is that we explicitly assume existence of
perturbation of the order $\lambda^3$, respectively $\eta
\lambda^2$, which is otherwise arbitrary. We take only the results
which are independent of potential perturbation. However, this is
a standard correct perturbational method. Such a treatment gives
us only finite number of conditions for Taylor coefficients. The
calculation is straightforward. Condition in the zeroth order
enables the calculation of
\[\rho^{(0)}_{12}=0;\quad \rho^{(0)}_{14}=0;\quad
\rho^{(0)}_{23}=0;\quad\rho^{(0)}_{34}=0,
\]
while in the second order\[ \rho^{(0)}_{13}=0;\quad
\rho^{(0)}_{24}=0;\quad \rho^{(2)}_{12}=0;\quad \rho^{(2)}_{14}=0;
\quad Im \rho^{(2)}_{23}=0;\quad Re
\rho^{(2)}_{23}=J(\rho^{(0)}_{33} - \rho^{(0)}_{22}),\]
\begin{equation}
\rho^{(0)}_{11}=\frac{\uG}{\dG}\rho^{(0)}_{22}; \quad \quad
\rho^{(0)}_{44}=\frac{\dG}{\uG}\rho^{(0)}_{33}.\label{anal}
\end{equation}

Here we clearly see the internal problem of the second order
approximation (\ref{approx2}). Neither the zeroth order of the
density matrix is resolved by stationary condition
(\ref{steady1}). We have still a two-dimensional subspace
(arbitrary $\rho^{(0)}_{22};\rho^{(0)}_{33}$) where the steady
state can be found (in the Liouville space of the density matrix).
The result (\ref{result2}) is the corollary of implicit assumption
of zero effect of higher order calculation, not justifiable from
the mathematical point of view. One can comprehend that including
a potentially higher order perturbation like (\ref{approx4}) will
define the zeroth order density matrix in space of our result
(\ref{anal}) with a high degree of arbitrariness.

The solution, which is correct to at least the zeroth order and
would have to be stable against perturbations, one must calculate
more precisely.

We notice that including the perturbation not as a potentiality,
but like a real effect changes the stability of the model. One can
prove in both interpretations that we obtain further condition
(once, in the order $\lambda^4$ or respectively $\eta \lambda^2 $)
\[\ug\rho^{(0)}_{22}=\dg\rho^{(0)}_{33}.
\] In the first interpretation we obtain in
the fourth order also further conditions; however, because of its
speculative character we will not publish it here. This
means that the situation, despite of its unpleasant character, 
is not hopeless. One can indicate instability and also the ways to
improve models are principally possible.  Let us notice that the
result obtained in this subsection is in accordance with all
previous calculations in $\lambda\rightarrow 0$ limit.

\section{The van Hove limit}
All our previous results were obtained in a way that is not just
standard in relaxation theory. We introduced perturbational scheme
(\ref{schema0}) for calculation of the second order kinetic
equation for model Hamiltonian (\ref{Ham}). We gave some physical
arguments for this choice. Nevertheless, the standard variant of
great popularity is of course the van Hove limit \cite{vHove}:
\begin{equation}
J \propto 1, \uG,\dG \propto \lambda^2. \label{vHove}
\end{equation}
We argue here that the problem with infinite time asymptotics of
the model (\ref{Ham}) (in the second order kinetic equation) is
reflected also in this well understood limit. To see this we
introduce " energetic" representation in eigenvectors of $H_S$
\begin{equation}
{\bf c_{II} }=\alpha c_2-\beta c_3,\quad {\bf c_{III}}=\alpha
c_3+\beta c_2
\end{equation}
where:
\[ \alpha=\frac{\sqrt{2}(1+\sqrt{1+\frac{4J^2}{\epsilon^2}})}{2\sqrt{\frac{4J^2}{\epsilon^2}+1+\sqrt{\frac{4J^2}{\epsilon^2}+1}}}\propto
1, \quad
\beta=\frac{\sqrt{2}J}{\epsilon\sqrt{\frac{4J^2}{\epsilon^2}+1+\sqrt{\frac{4J^2}{\epsilon^2}+1}}}\propto
\frac{J}{\epsilon}\]
 The model (\ref{Ham}) is now:
\[H= \epsilon c_1^{\dagger}c_1
+\frac{\epsilon}{2}(1+\sqrt{1+\frac{4J^2}{\epsilon^2}}){\bf
c_{III}^{\dagger}} {\bf c_{III}}
+\frac{\epsilon}{2}(1-\sqrt{1+\frac{4J^2}{\epsilon^2}}){\bf
c_{II}^{\dagger}}{\bf c_{II}}+ \sum_k \{\Omega_k B^{\dagger}_k
B_k+\]
\begin{equation}
G^{(1-2)}_k (B_k c_1^{\dagger}(\alpha {\bf c_{II}}+ \beta {\bf
c_{III}}) + B_k^{\dagger}(\alpha {\bf c_{II}^{\dagger}}+ \beta
{\bf c_{III}^{\dagger}})c_1) + G^{(3-4)}_k (B_k (\alpha {\bf
c_{III}^{\dagger} }-\beta{\bf c_{II}^{\dagger} })c_4 +
B_k^{\dagger}c_4^{\dagger}(\alpha {\bf c_{III} }-\beta{\bf c_{II}
})\} \label{vHam}
\end{equation}

We offer some comments regarding Hamiltonian (\ref{vHam}) and the
van Hove limit (\ref{vHove}). There are two bath induced channels
between levels 1,II and III,4 respectively, in analogy with the
previous treatment in site representation. What is the difference
is that there is no coherent transfer term in energetic
representation; on the other hand two weak bath induced channels
between levels 1,III and II,4 appeared. The strength of these
channels is proportional to $(JG)^2$, so this term is in the
second order kinetic theory of relevance in  the van Hove limit
only. (The region of physical applicability of (\ref{vHove}) does
not contain the regime specified before in connection with
(\ref{schema0}). We only clarify behavior of treated model
hamiltonian from another view.) These channels cause communication
between specified levels for short time regime, nevertheless both
the channels lie off the energy shell, so for the long time regime
this transfer is forbidden.
 Then the second order kinetic theory with integrated memory like
 \cite{ToMo} in infinite time
 forbids all the
communication between pair of levels 1+II and that of levels
III+4. Asymptotical stationary condition then has two linearly
independent solutions. Nevertheless
\[ c^{\dagger}_1 c_1 +{\bf c_{II}^{\dagger}}{\bf c_{II}}
\]
does not commute with the full Hamiltonian (\ref{vHam}), it is no
integral of motion. Consequently one can not use the long time
(Born-Markov) approximation upon looking for time asymptotics -
the result may depend also on short time transient effects.  The
result obtained in this way is also seemingly unstable against
higher order calculation. In Appendix B we give the complete
second order kinetic equation and its solution in the van Hove
scheme. The solution (\ref{vresult2}) of stationary condition
(\ref{steady1}) shows just the same asymptotic state of the
density matrix (and potential instability) like (\ref{anal}).

We stay here at quite real physical problem: Consideration whether
these levels are isolated and the transfer is strictly forbidden,
what suggests ordinary meaning of the energy conservation law, or
whether a limited value of electron density can be transferred.
The significance of the van Hove limit is here also questioned.

Last but not least, we may also think about more symmetrical case
of system - bath coupling of form: \[\sum_k \{G^{(1-2)}_k (B_k+
B^{\dagger}_k) (c_1^{\dagger}c_2 +  c_2^{\dagger}c_1) +
G^{(3-4)}_k (B_k + B^{\dagger}_k)( c_3^{\dagger}c_4 +
c_4^{\dagger}c_3).  \] The result for (\ref{schema0}) remains
unchanged. Then in energy representation also 1,III and II,4
channels appear that lie on energy shell for interaction with low
energy phonons, if these are present. Then the asymptotics in the
van Hove limit can be obtained as unique. However, the existence
of such phonons (and channels) is a serious change in physical
meaning of the entering model. Inapplicability of the van Hove
approach for physical regime $J<G^2$ is apparent.

\section{Physical consequences and conclusions}
Let us firstly discuss some physical consequences of possible
instability in the first scheme (\ref{schema1}). We have shown
that the result obtained in the second order may be unstable
against some correction arising in a higher order calculation. Let
us stay on the position that evolution generator
(Hamiltonian)(\ref{Ham}) is exact, and the only question is the
correctness of its study. Of course, in that case, our
construction of higher order correction is rather speculative, but
it proved the instability of the result. The full analysis of the
spectrum of the transfer matrix showed that the most `slow
excitation of the steady state' calculated in the second order has
its lifetime of the sixth order in $\lambda$. This gives that also
in the case where first correction to terms $W_{\{22\}\{33\}}$, $
W_{\{33\}\{22\}}$ is of the order six, it may cause such an
instability. This questions some standard results obtained in
everyday simulations. Unfortunately we have also some intuitive
physical arguments that one must really calculate at least six
order processes for achievement reliable result for transfer
between 2-3 sites.
 This is because the transfer rate really {\bf IS} a
process (at least) of the 6th order in specified scheme
(\ref{schema0}). Incoherent transfer term between sites 2 and 3
lies off energy-shell, consequently some collaboration with bath
modes for stabilization is necessary. All the Feynman graphs one
can draw must have for such a transfer great number of lines in
order to get on energetic sphere, implying high order of this
transfer. The reported delicate situation in the van Hove limit
also supported caution against straightforward use of kinetic
theory. Unfortunately, any higher order contribution calculation
is a difficult task, getting dramatically worse from order to
order. This fact will also in future cause the great popularity of
the " naive " treatment; in this direction, our expectation
concerning influence of our work is rather pessimistic. Greater
care about applicability of usual model methods is, on the other
hand and in the light of our results, more than appropriate.
Further investigation should be turned to higher order inspection
of master equation connected with Hamiltonian (\ref{Ham}) in
specified physical regime. Especially the question connected with
the infinite time asymptotics is a great challenge, of crucial
physical implication, and not satisfactorily resolved yet.
 \acknowledgements
 The present author is indebted to Prof. V. \v{C}\'{a}pek for his
 careful reading of the manuscript and valuable discussions concerning subject of this paper.
Support of grants 202/99/0182 of the Czech grant agency and
153/1999/B of the Grant agency of Charles University is gratefully
acknowledged.

\appendix
\section{Detailed analysis of spectrum}
Determination of the twelve eigenvalues does not meet with
problems, because these are the roots of quadratic polynoms. In
addition we would like mainly to know signs of real parts of the
eigenvalues, at least in the limit $\lambda \rightarrow 0$. Thus
we reduce complicated results into the Taylor series at least to
the order which gives the sign. :\[\xi_1=0, \quad \xi_2=-\uG-\dG\]
\[\xi_7=-i\frac{\epsilon}{2}-\frac{3\dG+\uG}{4} +i
\sqrt{\frac{\epsilon^2}{4}+i\epsilon\frac{\dG-\uG}{4}-\frac{(\dG-\uG)^2}{4}+J^2}\approx
-\dG
\]\[\xi_8=-i\frac{\epsilon}{2}-\frac{3\dG+\uG}{4} -i
\sqrt{\frac{\epsilon^2}{4}+i\epsilon\frac{\dG-\uG}{4}-\frac{(\dG-\uG)^2}{4}+J^2}\approx
-i\epsilon-\frac{\dG+\uG}{2}
\]
\[\xi_{11}=-i\frac{\epsilon}{2}-\frac{3\uG+\dG}{4} +i
\sqrt{\frac{\epsilon^2}{4}-i\epsilon\frac{\dG-\uG}{4}-\frac{(\dG-\uG)^2}{4}+J^2}\approx
-i\epsilon -\frac{\uG+\dG}{2}
\]\[\xi_{12}=-i\frac{\epsilon}{2}-\frac{3\uG+\dG}{4} -i
\sqrt{\frac{\epsilon^2}{4}-i\epsilon\frac{\dG-\uG}{4}-\frac{(\dG-\uG)^2}{4}+J^2}\approx
-\uG
\]
\[\xi_{15}= i \epsilon -\frac{\dG+\uG}{2}\]
\[
\xi_9=\xi_7^* ;\quad \xi_{10}=\xi_8^*; \quad \xi_{13}=\xi_{11}^*;
\quad \xi_{14}=\xi_{12}^*;\quad  \xi_{16}=\xi_{15}^*\]
 Further eigenvalues are roots of the 4-th order polynomial coming from the
 submatrix A. Though there is a formula which enables explicitly to
 extract the roots - so called Cardano formula, we do not use it
 because of its complicated form, and we only determine
 leading terms of the limit case $\lambda\rightarrow 0$ using the
 Taylor series. (This point provides no additional assumption about
 analytical structure of this dependence, all the results can be
 proved using  mean value theorem.)

\[\xi_3\approx -\uG-\dG ;
\quad \xi_4\approx \frac{-J^2(\uG+\dG)}{\epsilon^2}\]
\[\xi_5\approx
i\epsilon-\frac{\uG+\dG}{2} ;\quad\xi_6=\xi_5^*
\]

Notice: The complex square root used in formulae above is defined
into the upper half-plane of the complex plane (e.g. $Im \sqrt{}
\ge 0$ ).

\section{Time asymptotical solution of the second order kinetic equation of model in van Hove limit }
We start from (\ref{vHam}) and in the van Hove perturbational
scheme (\ref{vHove}). Organization of column vector of the density
matrix is following:
\[
\rho^T = (\begin{array}{ccccccccc}\rho_{11}, & \rho_{II,II}, &
\rho_{III,III}, & \rho_{44}, &  Re\rho_{II,III}, &
Im\rho_{II,III}, &
Re\rho_{1,II}, & Im\rho_{1,II},& Re \rho_{1,III},\end{array} \]
\[\begin{array}{ccccccc}
Im\rho_{1,III}, & Re\rho_{III,4}, & Im\rho_{III,4}, &
Re\rho_{II,4}, & Im\rho_{II,4}, & Re\rho_{14}, & Im\rho_{14}
\end{array} )\]
 Kinetic equations (\ref{kinetic}) obtained
here from e.g (\cite{ToMo}) are:
\renewcommand{\uG}{\Gamma_{\uparrow}^v}
\renewcommand{\dG}{\Gamma_{\downarrow}^v}
\begin{equation}
W^{(2)} =\left(\begin{array}{cccc}  A & 0 & 0 & 0\\ 0 & B & 0& 0
\\ 0 & 0 & C & 0 \\ 0 & 0 &  0 & D
\end{array}\right)
\label{vapprox2}
\end{equation}

where
\begin{equation}
A=\left(\begin{array}{cccccc} - \dG & \uG & 0& 0& \ab \uG & 0
\\ \dG & -\uG & 0 & 0 &\ab \dG & 0 \\ 0 & 0& -\dG & \uG &-\ab\uG& 0\\
0 & 0 &\dG & -\uG & -\ab \uG& 0 \\\frac{ \ab\dG}{2} &
-\frac{\ab\uG}{2} & \frac{\ab\dG}{2} & -\frac{\ab\uG}{2} &
-\frac{\uG + \dG}{2}& - \epsilon-2\Delta \\ 0 & 0 & 0 & 0 &
\epsilon+2\Delta &-\frac{\uG + \dG}{2}
\end{array}\right),
\end{equation}
\begin{equation}
B=\left(\begin{array}{cccc}-\frac{\uG + \dG }{2}& \epsilon +
\Delta & \frac{\ab\dG}{2} & 0 \\ -\epsilon -\Delta &-\frac{\uG + \dG }{2} & 0 & \frac{\ab\dG}{2} \\
-\frac{\ab\uG}{2} & 0 & -\dG
 & -\Delta \\ 0 & -\frac{\ab\uG}{2} & \Delta & -\dG
\end{array}\right),
\end{equation}
\begin{equation}
C=\left(\begin{array}{cccc}-\frac{\uG + \dG }{2}& \epsilon +\Delta
&  -\frac{\ab\uG}{2}& 0
\\ -\epsilon-\Delta &-\frac{\uG + \dG }{2} & 0 &  -\frac{\ab\uG}{2} \\  \frac{\ab\dG}{2}& 0 & -\uG
 & -\Delta\\ 0 &  -\frac{\ab\uG}{2} & +\Delta & -\uG
\end{array}\right),
\quad
D=\left(\begin{array}{cc}-\frac{\uG+\dG}{2} & \epsilon \\
-\epsilon & -\frac{\uG+\dG}{2}
\end{array}\right).
\end{equation}
where
:\[\Delta=\frac{\epsilon}{2}(\sqrt{1+\frac{4J^2}{\epsilon^2}}-1)\]
\[\ab=\frac{\beta}{\alpha}=\frac{2J}{\epsilon(1+\sqrt{1+\frac{4J^2}{\epsilon^2}})}\]
\[\uG=2\pi\alpha^2\sum_k [G^{(1-2)}_k]^2 \delta (\epsilon+\Delta-
\Omega_k)Tr_{Bath} \rho_{Bath} (B^{\dagger}_k B_k)\]\[ =2\pi
\alpha^2 \sum_k [G^{(3-4)}_k]^2 \delta (\epsilon+ \Delta-
\Omega_k)Tr_{Bath}( \rho_{Bath} B^{\dagger}_k B_k)\]
\[\dG=2\pi \alpha^2 \sum_k [G^{(1-2)}_k]^2
\delta (\epsilon+\Delta- \Omega_k)Tr_{Bath}( \rho_{Bath} B_k
B^{\dagger}_k)\]
\begin{equation}
 =2\pi\alpha^2\sum_k [G^{(3-4)}_k]^2 \delta (\epsilon+
 \Delta- \Omega_k)Tr_{Bath} (\rho_{Bath} B_k B^{\dagger}_k).
\end{equation}
One can verify that stationary condition (\ref{steady1}) is
satisfied by density matrix:
\begin{equation}\rho=C(\frac{\uG}{\dG+\uG} c_1^{\dagger}c_1
+\frac{\dG}{\dG+\uG}{\bf c_{II}^{\dagger}}{\bf
c_{II}})+(1-C)(\frac{\uG}{\dG+\uG}{\bf c_{III}^{\dagger}} {\bf
c_{III}} +\frac{\dG}{\dG+\uG}c_4^{\dagger}c_4)
\label{vresult2}\end{equation} with arbitrarily chosen constant
$C\in (0,1)$.

This proves the statement of main text.


\newpage

\begin{thebibliography}{99}
\bibitem{moje}{F. \v{S}anda} to be published elsewhere.
\bibitem{boltzmann}{R. Balescu,} Equilibrium and Nonequilibrium Statistical
Mechanics (J.Willey and Sons Inc., New York-London-Sydney-Toronto,
1975).
\bibitem{focker}{W. Brenig,} Statistical Theory of Heat:
Nonequilibrium Phenomena  (Springer-Verlag,Berlin-Heidelberg
1989).
\bibitem{pauli} {S.Fujita,} Introduction to Nonequilibrium Quantum Statistical Mechanics(W.B. Saunders Company,Philadelphia-London
1966).
\bibitem{Zwan}{R. Zwanzig,} Physica {\bf 30}(1964) 1109.
\bibitem{ToMo}{M. Tokuyama and H. Mori,} Progr. Theor. Phys. {\bf 55}
(1976) 411.
\bibitem{mark}{H. Bauer,} Probability Theory, (Walter de Gruyter,
Berlin-New York, 1996).
\bibitem{mark1}{V. \v{C}\'{a}pek,}Czech. J. Phys. 42 (1992) 317.
\bibitem{mark2}{R. Balescu,} Physica {\bf 27} (1961) 693.
\bibitem{mark3}{R.Swenson,}Physica {\bf 29} (1963) 1147.
\bibitem{Dav}{E. B. Davies,} Quantum Theory of Open Systems
(Academic Press, London, 1976).
\bibitem{vHove}{L. van Hove,} Physica {\bf 21} (1955) 517.; ibid. {\bf
23} (1957) 441.
\bibitem{capbar}{V. \v{C}\'{a}pek, I. Barv\'{\i}k,}
Physica A {\bf 294} (2001) 388.
\bibitem{HSR1}{H. Haken, G. Strobl, } "Exact Treatment of Coherent
and Incoherent Triplet Exciton Migration" in {\it The  Triplet
State},ed. by A.Zahlan (Cambridge Univ. Press, London , 1967) p.
311-314.
\bibitem{HSR2}{P. Reineker, H. Haken} "The Coupled Coherent and
Incoherent Motion of Frenkel Exciton in Molecular Crystals" in
Localization and Delocalization in Quantum Chemistry, Vol. II ed.
by O.Chalvet, R. Daudel, R. Diner, P. Malrieu (Reidel, Dortrecht,
Boston 1976) pp. 185.-194.
\bibitem{cap4}{V. \v{C}\'{a}pek,}
http://arxiv.org/abs/cond-mat/0012056,Europ. Phys. Journal B
(2002) - in press. \end{thebibliography}
\end{document}